\documentclass[10pt, letterpaper]{IEEEtran}
\IEEEoverridecommandlockouts
\usepackage{cite}
\usepackage{amsmath,amssymb,amsthm}
\usepackage{bbm}
\usepackage{algpseudocode}
\usepackage{graphicx}
\usepackage{textcomp}
\usepackage{mathtools}
\usepackage{xcolor}
\usepackage{algorithm}
\usepackage{bm}
\usepackage{booktabs}

\def\BibTeX{{\rm B\kern-.05em{\sc i\kern-.025em b}\kern-.08em
    T\kern-.1667em\lower.7ex\hbox{E}\kern-.125emX}}
\usepackage{url}

\usepackage{tikz}
    \usetikzlibrary{shapes.arrows}
    
\usepackage[caption=false]{subfig}
\usepackage{pgfplots}
\pgfplotsset{
compat=1.3,
legend style={font=\footnotesize, fill opacity=0.7,  draw opacity=1, text opacity=1, draw=white!15!black, legend cell align=left, align=left}, 
width=6cm, 
height=6cm,
yminorticks=false,
xminorticks=false,
title style={font=\small},
tick style={color=black},
tick label style={font=\small},
grid style={line width=.1pt, draw=gray!20},
major grid style={line width=.1pt,draw=gray!20},
}
\usepgfplotslibrary{colorbrewer}
\usepgfplotslibrary{groupplots}
\usepgfplotslibrary{fillbetween}
\usetikzlibrary{decorations.pathreplacing}
\usetikzlibrary{plotmarks}
\usepackage[]{fancyhdr}
\usetikzlibrary{patterns}
\pgfplotsset{every tick label/.append style={font=\footnotesize}}

\usepackage{tikz}
\usetikzlibrary{shapes, arrows, positioning}

\usepackage{glossaries}
\newacronym{goc}{GoC}{Goal-oriented Communication}
\newacronym{ps}{PS}{Periodic Scheduling}
\newacronym{pp}{PP}{Periodic Policy}
\newacronym{iot}{IoT}{Internet of Things}
\newacronym{ml}{ML}{machine learning}
\newacronym{dl}{DL}{Deep Learning}
\newacronym{marl}{MARL}{multi-agent reinforcement learning}
\newacronym{rl}{RL}{reinforcement learning}
\newacronym{decpomdp}{Dec-POMDP}{Decentralized Partially Observable Markov Decision Process }
\newacronym{icn}{ICN}{Information-Centric Networking}
\newacronym{jscc}{JSCC}{Joint Source-Channel Coding}
\newacronym{acnomdp}{ACNO-MDP}{Action-Contingent Noiselessly Observable MDP}
\newacronym{pomdp}{POMDP}{Partially Observable Markov Decision Process}
\newacronym{lidar}{LIDAR}{Laser Imaging Detection and Ranging}
\newacronym{uav}{UAV}{Unmanned Aerial Vehicle}
\newacronym{dqn}{DQN}{Deep Q-Network}
\newacronym{dnn}{DNN}{deep neural network}
\newacronym{dial}{DIAL}{Differentiable Inter-Agent Learning}
\newacronym{api}{API}{Alternate Policy Iteration}
\newacronym{mdp}{MDP}{Markov Decision Process}
\newacronym{fov}{FoV}{Field of View}
\newacronym{cnn}{CNN}{Convolutional Neural Network}
\newacronym{nn}{NN}{neural network}
\newacronym{ddql}{DDQL}{Distributed Deep Q-Learning}
\newacronym{pdf}{PDF}{Probability Density Function}
\newacronym{ndpomdp}{ND-POMDP}{Networked Distributed Partially Observable Markov Decision Process}
\newacronym{ncs}{NCS}{Networked Control System}
\newacronym{cps}{CPS}{Cyber-Physical System}
\newacronym{radam}{RAdam}{Rectified Adam}
\newacronym{cdf}{CDF}{cumulative distribution function}
\newacronym{mpc}{MPC}{Model Predictive Control}
\newacronym{rv}{rv}{Random Variable}
\newacronym{qoe}{QoE}{Quality of Experience}
\newacronym{tlc}{TLC}{Telecommunications}
\newacronym{cml}{CML}{communications for machine learning}
\newacronym{mlc}{MLC}{machine learning for communications}
\newacronym{drl}{DRL}{Deep Reinforcement Learning}
\newacronym{rf}{RF}{Radio Frequency}
\newacronym{urllc}{URLLC}{Ultra-Reliable and Low-Latency Communications}
\newacronym{fl}{FL}{federated learning}
\newacronym{kpi}{KPI}{Key Performance Indicators}
\newacronym{mec}{MEC}{Mobile Edge Computing}
\newacronym{ei}{EI}{Edge Intelligence}
\newacronym{bs}{BS}{base station}
\newacronym{sdn}{SDN}{Software Defined Networking}
\newacronym{mimo}{MIMO}{Multiple-Input Multiple-Output}
\newacronym{gp}{GP}{Gaussian Process}
\newacronym{iiot}{IIoT}{Industrial Internet of Things}
\newacronym{csi}{CSI}{Channel State Information}
\newacronym{sgd}{SGD}{Stochastic Gradient Descent}
\newacronym{iid}{i.i.d.}{independent and identically distributed}
\newacronym{ofdm}{OFDM}{Orthogonal Frequency Division Multiplexing}
\newacronym{los}{LOS}{Line-of-Sight}
\newacronym{nlos}{NLOS}{Non-Line-of-Sight}
\newacronym{snr}{SNR}{Signal to Noise Ratio}
\newacronym{rb}{RB}{Resource Block}
\newacronym{6g}{6G}{sixth generation}
\newacronym{ai}{AI}{artificial intelligence}
\newacronym{sfl}{SFL}{Synchronous Federated Learning}
\newacronym{frfl}{FRFL}{Fixed Rate Federated Learning}
\newacronym{pgm}{PGM}{Probabilistic Graphical Model}
\newacronym{hmm}{HMM}{Hidden Markov Model}
\newacronym{elbo}{ELBO}{Evidence Lower Bound}
\newacronym{pmf}{PMF}{Probability Mass Function}
\newacronym{smab}{SMAB}{Stochastic Multi-Armed Bandit}
\newacronym{mab}{MAB}{multi-armed bandit}
\newacronym{mc}{MC}{Monte Carlo}
\newacronym{is}{IS}{Importance Sampling}
\newacronym{dms}{DMS}{discrete memoryless source}
\newacronym{ucb}{UCB}{upper confidence bound}
\newacronym{ser}{SER}{Symbol Error Rate}
\newacronym{sc}{SC}{Semantic Communications}
\newacronym{voi}{VoI}{Value of Information}
\newacronym{nlp}{NLP}{natural language processing}
\newacronym{ts}{TS}{Thompson Sampling}
\newacronym{cmab}{CMAB}{contextual multi-armed bandit}
\newacronym{rccmab}{RC-CMAB}{rate-constrained \gls{cmab}}
\newacronym{rcmab}{R-CMAB}{remote \gls{cmab}}
\newacronym{ib}{IB}{information bottleneck}
\newacronym{merl}{MERL}{maximum entropy reinforcement learning}
\newacronym{pac}{PAC}{Probably Approximately Correct}
\newacronym{aoi}{AoI}{Age of Information}
\newacronym{aoii}{AoII}{Age of Incorrect Information}
\newacronym{uoi}{UoI}{Urgency of Information}
\newacronym{vqvae}{VQ-VAE}{Vector Quantized Variational Autoencoder}
\newacronym{vae}{VAE}{Variational Autoencoder}
\newacronym{psnr}{PSNR}{Peak Signal to Noise Ratio}
\newacronym{mse}{MSE}{Mean Square Error}
\newacronym{lstm}{LSTM}{Long Short-Term Memory}
\newacronym{iomdp}{IOMDP}{Intermittently Observable Markov Decision Process}
\newacronym{pi}{PI}{Policy Iteration}
\newacronym{mpi}{MPI}{Modified Policy Iteration}
\newacronym{iql}{IQL}{Intermittent Q-Learning}
\newacronym{nc}{NC}{Network Core}
\newacronym{pql}{PQL}{Peekaboo Q-Learning}
\newacronym{eff_com}{EC}{Effective Communication}
\newacronym{ibr}{IBR}{Iterated Best Response}
\newacronym{ne}{NE}{Nash Equilibrium}
\newacronym{map}{MAP}{maximum a posteriori}
\newacronym{ade}{ADE}{Alternating Defense from Eavesdropping}
\newacronym{pde}{PDE}{Packing Defense from Eavesdropping}
\newacronym{oposg}{OPOSG}{one-sided partially observable stochastic game}

\DeclareMathOperator{\Mod}{mod}
\newcommand{\E}[1]{\mathbb{E}\left[ #1 \right]} % expectation
\newcommand{\mc}[1]{\mathcal{#1}}   % mathcal abbreviation
\newcommand{\mb}[1]{\mathbf{#1}}    % mathbf abbreviation
\DeclareMathOperator*{\argmax}{arg\,max}    % argmax
\algtext*{EndWhile}% Remove "end while" text
\algtext*{EndIf}% Remove "end if" text
\algtext*{EndFor}% Remove "end for" text

\def \fheight {0.36\columnwidth}

\def \sfwidth{0.95\columnwidth}
\def \sfheight {0.5\columnwidth}
\def \boxside {0.3\columnwidth}

\def\boxheight{0.24\columnwidth}

\definecolor{color0}{HTML}{00429D}
\definecolor{color1}{HTML}{844D99}
\definecolor{color2}{HTML}{C3608E}
\definecolor{color3}{HTML}{EF8078}
\definecolor{color4}{HTML}{FFB047}
\definecolor{darkslategray38}{RGB}{38,38,38}
\definecolor{darkblue}{HTML}{00429D}
\definecolor{darkgreen}{HTML}{005c00}
\definecolor{gold}{HTML}{D4AF37}
\definecolor{darkred}{HTML}{910000}

\newtheorem{theorem}{Theorem}

\makeatletter
\title{Secure Goal-Oriented Communication:\\ Defending against Eavesdropping Timing Attacks}

\author{Federico Mason, \IEEEmembership{Member, IEEE}, Federico Chiariotti, \IEEEmembership{Senior Member, IEEE},\\Pietro Talli, \IEEEmembership{Graduate Student Member, IEEE}, and Andrea Zanella, \IEEEmembership{Senior Member, IEEE}
\thanks{The authors (emails: federico.mason@unipd.it, federico.chiariotti@unipd.it, pietro.talli@phd.unipd.it, andrea.zanella@unipd.it) are with the Department of Information Engineering, University of Padova, Via G. Gradenigo 6/B, 35131, Padova, Italy. This project was funded under the National Recovery and Resilience Plan (NRRP), funded by the European Union NextGenerationEU Project as part of the ``RESTART'' partnership (PE0000001).}}

\begin{document}

\maketitle

\begin{abstract}
\gls{goc} is a new paradigm that plans data transmission to occur only when it is instrumental for the receiver to achieve a certain goal.
This leads to the advantage of reducing the frequency of transmissions significantly while maintaining adherence to the receiver's objectives.
However, \gls{goc} scheduling also opens a timing-based side channel that an eavesdropper can exploit to obtain information about the state of the system.
This type of attack sidesteps even information-theoretic security, as it exploits the timing of updates rather than their content.
In this work, we study such an eavesdropping attack against pull-based goal-oriented scheduling for remote monitoring and control of Markov processes.
We provide a theoretical framework for defining the effectiveness of the attack and propose possible countermeasures, including two practical heuristics that provide a balance between the performance gains offered by \gls{goc} and the amount of leaked information.
Our results show that, while a naive goal-oriented scheduler allows the eavesdropper to correctly guess the system state about $60$\% of the time, our heuristic defenses can halve the leakage with a marginal reduction of the benefits of goal-oriented approaches.
\end{abstract}


\glsresetall

\section{Introduction}
\label{sec:intro}

\begin{tikzpicture}[remember picture, overlay]
      \node[draw,minimum width=4in] at ([yshift=-1cm]current page.north)  {This manuscript has been submitted to IEEE Journal on Selected Areas in Communication.};
\end{tikzpicture}

Over the past few years, the \gls{goc} paradigm has attracted a significant amount of interest from the research community. The concept was advanced by Warren Weaver in his 1949 introduction to Shannon's theory of communication~\cite{shannon1949mathematical}, and regards the design of more advanced communication protocols that go beyond the mere transmission of bits and consider the meaning and usefulness of the data for the receiver in the decision over what and when to transmit. 
On the other hand, a practical implementation of these ideas requires powerful machine learning techniques~\cite{gunduz2022beyond} and, therefore, has only recently become feasible.

Goal-oriented approaches were initially applied to compression~\cite{deepjscoding} and have successively been extended to scheduling strategies that consider contextual and past information~\cite{fountoulakis2023goal}.
These initial studies have shown that \gls{goc} leads to impressive performance advantages, fostering research on more practical aspects including security against eavesdropping attacks~\cite{guo2024survey}.
The most common approach to enhance \gls{goc} security is to train the transmitter to encrypt the data~\cite{tung2023deep}, modifying the encoding mechanism to trigger an incorrect semantic interpretation by possible eavesdroppers~\cite{liu2023semprotector}, while allowing the intended receiver to decode the original message.
In this regard, information-theoretic approaches~\cite{kung2018compressive} can provide more solid confidentiality guarantees~\cite{chen2024nearly}, but only under specific assumptions on the nature of the encoder and decoder.

Although the above mechanisms address the risk of leaking information through the \textit{content} of the transmitted data, another specific vulnerability of \gls{goc} systems has been mostly neglected so far: side-channel attacks that aim to infer the state of the system from the \textit{timing} of messages~\cite{van2015clock}.
This is particularly critical for \gls{iot} applications or other resource-constrained monitoring systems, where \gls{goc} is used to reduce the frequency of updates according to the status of the monitored process.
In these scenarios, timing attacks can leak information about the content of transmitted packets (i.e., the state of the process) even when using one-time pad encryption or information-theoretic security.

In this work, we analyze the secrecy of a goal-oriented scheduling system under a timing attack from an eavesdropper.
Specifically, we consider a pull-based communication scenario in which a controller node maintains an online estimate of the state of a remote Markov process, in order to monitor or control the process itself~\cite{talli2024pragmatic}.
The state of the process is not directly observable by the controller node but is continuously tracked by a sensor node that can transmit the current state to the other node upon request.
The goal of the controller is to schedule status update transmissions from the sensor node to obtain high reward for its local task, while minimizing the channel occupancy and limiting the information that can be inferred by an eavesdropper from the timing between consecutive transmissions in either direction.

%We show that adapting the scheduling policy to balance cost with task's performance reduces the opacity of the system by opening a side channel that an eavesdropper can use to gain information. 
%In other words, by monitoring the distribution of inter-transmission intervals, it is possible to estimate the current and past states of the Markov process.
%In fact, by knowing the state-transition probability matrix of the Markov Process, an eavesdropper can possibly estimate the current and past states of the Markov process by monitoring the distribution of inter-transmission intervals.

We consider the critical condition in which the eavesdropper knows the state-transition probability matrix of the monitored Markov process and the policy used by the controller to schedule transmissions from the sensor node. In addition, the eavesdropper knows the timing of all past transmissions.
Hence, we analyze the trade-off between secrecy, which depends on the information leakage of the system, and performance, measured in terms of reward for the controller node and transmission efficiency.
We consider four different strategies: a pure goal-oriented approach, which optimizes performance disregarding security aspects; a periodic scheduling that prevents timing attacks, but loses \gls{goc} advantages; and two novel heuristics that reduce information leakage while preserving performance.
The analysis is repeated both in a monitoring scenario, where the controller only aims at estimating the status of the process, and in a control scenario, where the system's evolution can be altered by the controller itself. 

To our knowledge, this manuscript is the first to consider the secrecy implications of timing attacks against \gls{goc}, and includes the following main contributions:
\begin{itemize}
    \item We provide a rigorous model of timing attacks in \gls{goc}, defining information leakage as a function of the time for which confidentiality must be ensured.
    \item We prove that finding a game-theoretical equilibrium when both the legitimate agent and the eavesdropper are rational actors is a computationally hard problem.
    \item We propose a heuristic algorithm, named \gls{ade}, which allows the legitimate agent to compute the information leakage in real time and take countermeasures accordingly.
    \item We propose a lighter heuristic algorithm, named \gls{pde}, which pursues the same objective as \gls{ade}, but with a lower complexity, enabling its implementation as a look-up table.
    \item We evaluated the effectiveness of timing attacks and defensive strategies through Monte Carlo simulations for both estimation and control scenarios.
\end{itemize}
A preliminary version of this work was presented as a conference paper in~\cite{mason2025timing}.
This manuscript extends our previous results by introducing the \gls{pde} policy and analyzing the overall framework in the case of control applications.

The remainder of the paper is organized as follows.
First, Sec.~\ref{sec:sota} reviews state-of-the-art security schemes in semantic and \gls{goc} communication.
Hence, Sec.~\ref{sec:model} presents the \gls{goc} model, drawing from the results of our previous work~\cite{talli2024pragmatic}, while Sec.~\ref{sec:attack} presents the eavesdropping attack and the game-theoretical framework.
Subsequently, Sec.~\ref{sec:defense} introduces the heuristic algorithms to mitigate information leakage in the system, and Sec.~\ref{sec:results} discusses our simulation settings and results.
Finally, Sec.~\ref{sec:conc} concludes the article and describes possible avenues for future research.

\section{Related Work}
\label{sec:sota}
As \gls{goc} is still a relatively new paradigm, research on its security aspects, such as eavesdropping attacks, is still in its infancy. The existing \gls{goc} security literature mostly focuses on a subclass of \gls{goc} problems which focuses on reconstructing the transmitted information directly, without any memory or time-dependence. In this context, timing attacks are not meaningful, and the focus is on the content of each message.

Besides an early work using an information bottleneck approach~\cite{kung2018compressive}, previous studies mainly deal with eavesdropping attacks using deep learning~\cite{guo2024survey}. More recently, the authors of \cite{chen2024nearly} provide a near-information-theoretic security approach for semantic communication. The authors adopt the classic approach in information-theoretic security by considering a legitimate receiver with a higher \gls{snr} than the eavesdropper, allowing the semantic scheme to exploit this advantage by properly encoding the semantic symbols.
A very common semantic communication approach is deep \gls{jscc}. This model was adapted to include Shannon secrecy in~\cite{tung2023deep}, extending the information-theoretic approach to learning-based semantic encoders, whose constellations are learned rather than hand-designed: in this case, the learning algorithm converges to a secret semantic encoding by using secrecy as an additional objective function, exploiting similar principles as traditional information-theoretic security.
Interestingly, the \gls{jscc} protection module can be implemented after semantic encoding (\cite{tung2023deep}), before encoding (\cite{xu2023deep}), or integrated within the encoder (\cite{li2024secure,shi2025secure}), with similar results and trade-offs in terms of secrecy and image transfer quality.

Another example of semantic encryption is given in~\cite{liu2023semprotector}, where eavesdroppers adopt a model inversion approach to retrieve the original information.
The use of explicit semantic features of the image~\cite{rong2025semantic} can also be used to generate shared secrets between the transmitter and the legitimate receiver that can be used to improve security. The same concept has been extended to the vision transformer architecture in~\cite{huang2025secure}.
Finally, the authors of~\cite{tang2025towards} adopt steganographic techniques to fool the eavesdropper into recovering an unrelated image, while keeping the meaningful content secure.

Active attacks that go beyond eavesdropping have been designed and tested against semantic communication in~\cite{xu2024csba}, whose authors consider the integrity of messages and the reliability of the application as dual objectives.
More complete threat models for semantic communication are given in~\cite{shen2024secure, yang2024secure}, which include attacks against various components of the system, including the training process.
%the literature on semantic communication security, which was thoroughly reviewed in~\cite{meng2025survey},
We observe that these previous works focus on securing the content of the current semantic message, without considering previous transmissions~\cite{meng2025survey}.
In addition, side-channel attacks, such as the one considered in this work, have been mostly neglected by the semantic communication literature. This is a critical issue, because this type of attack can be effective even when the content of messages is perfectly secure (e.g., when protected through one-time padding).

Interestingly, side-channel attacks have been considered in other practical scenarios, such as cloud scheduling.
For example, the work in~\cite{kadloor2015mitigating} analyzes a model in which a scheduler dispatches computing jobs to servers to satisfy clients with different arrival times. 
In this scenario, a malicious entity can infer the traffic patterns of legitimate users by measuring the scheduler's response time. 
A possible defense is the partial randomization of task execution times~\cite{yoon2016taskshuffler}, which significantly reduces information leakage through the side channel at the cost of lower system efficiency. 
Similar considerations were applied to the field of \gls{icn}, in which caching is used to infer information about user requests and the popularity of content~\cite{mohaisen2014timing}.

\begin{table*}
\scriptsize
\caption{Model notation.}
\label{tab:model}
\centering
\begin{tabular}{ll|ll|ll|ll}
\toprule
Symbol & Description & Symbol & Description & Symbol & Description & Symbol & Description \\
\midrule
$\mc{S}$ & State space & $\mc{A}$ & Action Space & $\bm{P}$ & Transition probability matrix & $\gamma$ & Discount factor \\
$R(\cdot)$ & Total reward function & $r_B(\cdot)$ & Bob's reward function & $r_A(\cdot)$ & Alice's reward function & $T_{\text{max}}$ & Maximum timing signal \\
$\beta$ & Transmission cost & $\bm{\mu}$ & Steady-state distribution & $\psi(\cdot)$ & Communication policy & $\pi(\cdot)$ & Control policy \\
$D$ & Opacity time gap & $L_E(n;D)$ & Information leakage & $\bm{\phi}_E$ & Eve's belief distribution & $L_{\text{min}}$ & Minimum leakage \\
$\eta$ & Eve's estimate & $\sigma(\cdot)$ & Communication policy & $f_k(s)$ & Forward probability & $b_k(s; n)$ & Backward probability \\
$L_{\text{low}}$ & ADE's lower threshold & $L_{\text{high}}$ & ADE's higher threshold & $\xi_{\sigma}^{(s^*,\tau)} (\cdot)$ & Single deviation policy & $H^*$ & PDE's target entropy \\
$\theta$ & Density decay & $H(\cdot)$ & Entropy function & $\zeta_{\tau,s}(s')$ & $\tau$-step transition probability & $\delta(\cdot)$ & Kronecker delta function \\
\bottomrule
\vspace{-0.3cm}
\end{tabular}
\end{table*}

Finally, we consider related work from another field, namely, remote estimation and control: studies from this area are not closely related to semantic communication and \gls{goc}, but they approach similar problems from another angle, and some of their conclusions can be applied to the scenarios studied in this manuscript.
In the case of a remote estimation scenario, the secrecy of monitoring systems against side channel attacks is closely related to the concept of \emph{opacity}.
In the estimation literature, a system is considered opaque if an eavesdropper with limited observations is unable to estimate some restricted information~\cite{mazare2004decidability}, including the identity of a client or whether the system enters a set of secret states.
The analysis of opacity has been extended to $K$-step observations~\cite{yin2017new} and even scenarios in which the eavesdropper has access to the entire observation history~\cite{saboori2011verification}.
In information-theoretic terms, opacity can be defined as the difference between the entropy of the belief distribution of the legitimate monitor and that of the eavesdropper~\cite{chen2017quantification}.

In control scenarios, where the legitimate agent can affect the state evolution of the system through actions, but the control policy is known to the eavesdropper, opacity is more difficult to achieve, and its formal verification becomes a highly complex~\cite{liu2020notion} or even undecidable problem~\cite{berard2015probabilistic}.
At the same time, the ability to affect the state of the system enables agents to actively improve security by inserting fictitious events~\cite{ji2018enforcement} to confuse eavesdroppers. 
This inherent complexity makes it critical to design \gls{goc} policies that optimize control performance under opacity constraints, or optimize both simultaneously.
To the best of our knowledge, the current literature considers only the problem of maximizing the opacity of the initial state or the current state. In this work, we generalize the problem considering the opacity of the entire system history, which is a significantly more challenging problem.

\section{Goal-Oriented Communication Model}
\label{sec:model}

We consider a remote control scenario in which one node (Alice) can instantaneously observe the state of a discrete-time Markov chain defined by a state space $\mc{S}$ and a transition matrix $\mb{P}$.
We denote by $s(n) \in \mc{S}$ the state of the process at time step $n$ and by $\bm{\mu}_0$ the initial probability distribution of the state.
A second node (Bob) is assigned the task of controlling or estimating the process (depending on the scenario considered) by choosing an action over a state space $\mc{A}$. Both Alice and Bob have complete knowledge of $\mb{P}$ and $\bm{\mu}_0$, but Bob cannot observe $s(n)$ directly and must rely on Alice's transmissions to update his information about the current process state. 
The notation used in our model is reported in Tab.~\ref{tab:model}.

We consider a \emph{pull-based} configuration in which, at each time step $n$, Bob must decide whether to ask Alice for an update, thus incurring a communication cost $\beta \in \mathbb{R}^+$, or to estimate the current state of the Markov chain from the information he already knows.
We denote Bob's binary communication decision as $c(n) \in \{0, 1\}$, with $c(n)=1$ in the case of transmission, and $c(n)=0$ otherwise. Moreover, we assume a maximum number of steps, $T_{\max}$, after which Bob always requests an update. This parameter is necessary for the tractability of the analysis, but its impact can be arbitrarily minimized by considering large values of $T_{\max}$.

We then define a \textit{task reward function} $r_B:\mc{S} \times \mc{A} \rightarrow \mathbb{R}$ that determines the performance of Bob's task (estimation or control).
We remark that, when considering remote estimation scenarios, Bob's action consists of estimating the state from the available information, that is, $a(n)=\hat{s}(n)$. The action space then corresponds to the state space, and the transition probabilities of the Markov process are independent of the selected action, i.e., $P(s'|s,a)=P(s'|s,a')\ \forall$~$a,a'\in\mc{A}$.
Hence, the task reward function is equal to $1$ if the state estimate matches the actual state, and $0$ otherwise, i.e., $r_B(s,\hat{s})=\delta(s,\hat{s})$, where $\delta(\cdot,\cdot)$ is the Kronecker delta function. %, equal to $1$ if the two arguments are the same and $0$ otherwise.

We also introduce the \textit{communication reward function} $r_A:\{0,1\}\to\mathbb{R}$, with $r_A(c)=-\beta c$, where $\beta$ is a communication cost that is paid only when Bob asks for a transmission ($c=1$). The total reward is then given by the combination of the task reward and the (negative or null) communication reward:
\begin{equation}
    R(s,c,a)=r_B(s,a)+r_A(c).
    \label{eq:effective_reward}
\end{equation}

Therefore, Bob's objective is to find the communication policy that maximizes the expected cumulative reward
\begin{equation}
    G(n)=E\left[ \sum_{k=n}^{+\infty} \gamma^{(k-n)} R(s(k),c(k),a(k)) \right],
\label{eq:return_effective}
\end{equation}
where $\gamma \in [0,1)$ is the exponential discount factor.
The described problem is a remote \gls{pomdp}, comprehensively characterized by the tuple $\langle \mc{S}, \mc{A}, \mb{P}, r_B(\cdot), \gamma, T_{\max}, \beta \rangle$.

We assume that the communication delay is shorter than the time step of the underlying Markov process, so that when Alice transmits, Bob receives the state information instantaneously (i.e., within the same time slot). Using the state updates from Alice and his knowledge of $\mb{P}$, Bob keeps a local estimate of the state probability distribution of the remote process, that is to say, a \textit{belief} on the process state that we denote as $\bm{\zeta}$.

Let $\zeta_{\Delta,s}(s')$ represent Bob's estimate of the probability that the process will be in state $s'$ in $\Delta$ steps, given that Alice just reported that the process was in state $s$. This probability can be computed recursively as
\begin{equation}
  \zeta_{\Delta,s}(s')=\sum_{s''\in\mc{S}}P(s'|s'';\pi)\zeta_{\Delta-1,s}(s'')\,,
  \label{eq:zeta}
\end{equation}
with $\zeta_{0,s}(s')=\delta(s,s')$. Bob's control policy $\pi$ is a parameter of the transition probabilities because, in the control scenario, the evolution of the Markov process is generally affected by Bob's actions. In the estimation case, we can simplify \eqref{eq:zeta} to $\zeta_{\Delta,s}(s')=\mb{P}^{\Delta}(s,s')$, i.e., to the element with indices $s$ and $s'$ of the $\Delta$-th power of the transition matrix, as the evolution of the system does not depend on the control policy $\pi$.

Since each transmission represents a renewal of Bob's beliefs, the current estimate of the process state can be summarized by the last received state $s$ and the time $\Delta$ since the last update~\cite{talli2024pragmatic}. 
Therefore, Bob's optimal decisions depend only on $(s,\Delta)$, which reduces the complexity of the problem.
Importantly, the \gls{mpi} scheme given in~\cite[Alg. 1]{talli2024pragmatic} can find the jointly optimal \gls{goc} policy $\psi: \mc{S} \times \mathbb{Z}^+ \rightarrow \{0, 1\}$ and the associated control policy $\pi: \mc{S} \times \mathbb{Z}^+ \rightarrow \mc{A}$, in polynomial time with respect to the size $|\mc{S}|$ of the state space.

Practically, any time Bob receives a state update $s\in S$ from Alice, he can determine his future control actions in advance, as well as the optimal number of time steps to wait before the next update request, which is the smallest $\Delta$ such that $\psi(s,\Delta)=1$.
We denote the transmission request scheduling function as $\sigma: \mc{S} \rightarrow \mathbb{Z}^+$, defined as
\begin{equation}\label{eq:timing_policy}
    \sigma(s)=\inf\{\Delta \in \mathbb{N} : \psi(s, \Delta)=1\}\,.
\end{equation}
This function then determines the inter-transmission intervals. 

\section{Eavesdropping Attack}
\label{sec:attack}
\begin{figure}[t!]
\input{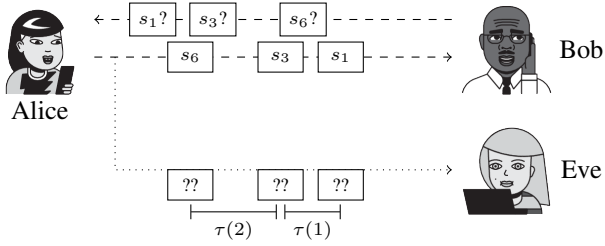}
\caption{The goal-oriented eavesdropping attack: Eve cannot decipher Alice's updates, but the timing signal $\tau$ allows her to estimate the state of the remote Markov process.}
\label{fig:diagram}
\end{figure}

We assume that an eavesdropper (Eve) knows the Markov process statistics represented by $\mb{P}$ and $\bm{\mu}_0$, and the transmission request scheduling $\sigma(\cdot)$.
%The latter can be determined in the same manner as for Bob. 
However, Eve cannot directly observe the process, nor read the content of Alice's transmissions. Therefore, she tries to gain information about the state of the Markov chain by observing the intervals between consecutive Bob's requests.  
From Eve's perspective, the system is a \gls{hmm}, where the timing signals $\tau$, i.e., the intervals between consecutive state updates, are the observations from which she determines the \gls{map} estimate of the Markov source state.
A scheme of the overall scenario is reported in Fig.~\ref{fig:diagram}.

\subsection{Information Leakage}

We now consider the problem of preventing Eve from acquiring information about the remote process state from Alice and Bob's communication.\footnote{We specify that our formulation does not consider the initial knowledge of Eve over the Markov source. If the initial state distribution is low-entropy, the mixing time of the chain might be quite long, which leads to an edge case whose analysis is left to future work.}
We define the secrecy objective using the concept of the \textit{opacity time gap}, denoted as $D$.
This gap represents the number of past time steps for which information on the Markov chain should remain undisclosed. 
Let $\bm{\phi}_E(n;d)$ denote Eve's belief distribution about the process state at time $n-d$, given that she has listened to the channel up to time $n$.
Therefore, $[\bm{\phi}_E(n;d)](s)$ is the probability that $s(n-d)=s \in S$ from Eve's perspective. 

We define the \textit{information leakage} at time step $n$ as
\begin{equation}
    L_E(n;D)=\max_{d\in\{0,\ldots,D\}} \left\{1-\frac{H\left(\bm{\phi}_E(n;d)\right)}{H_0} \right\},
\label{eq:leakage}
\end{equation}
where $H(\cdot)$ is the Shannon information theoretic entropy defined as $H(\mb{p})=-\sum_{s\in\mc{S}}p(s)\log_2(p(s))$, with $p(\cdot)$ denoting the probability distribution of the process state~\cite{shannon1949mathematical}. The denominator $H_0 = \log_2(|\mc{S}|)$ is a normalization constant. We note that $L_E(n;D)=0$ only if $\bm{\phi}_E(n;d)$ is uniform in the state space for any delay $d \in \{0,\ldots,D\}$, which means that Eve does not have information on the state of the system in the last $D$ steps. On the other hand, $L_E(n;D)=1$ if $\bm{\phi}_E(n;D)=\delta(s,s_{n-d})$ for some $d\leq D$, i.e., Eve has perfect knowledge of the state of the system in at least one of the last $D$ steps.
%As discussed in more detail in Sec.~\ref{ssec:ade}, a periodic scheduling policy always minimizes the leakage, as the timing of state updates becomes independent of the monitored process state. 
On the other hand, Eve is always able to determine the steady-state distribution $\bm{\mu}$ of the system.
This implies that the leakage can never be less than
%$L_E(n;D)$ tends to zero only when $\bm{\mu}$ is a uniform distribution, while, in the general scenario, 
%the minimum leakage is
\begin{equation}
    L_{\text{min}}= 1-\left(H_0\right)^{-1}H\left(\bm{\mu}\right)\geq 0\;.
\label{eq:leak_min}
\end{equation}
Therefore, zero leakage can be achieved only for processes with a uniform steady-state distribution $\bm{\mu}$, for which $H\left(\bm{\mu}\right)=H_0$, whereas $L_{\text{min}}>0$ in the general case. 
Finally, we observe that Eve's best estimate of state $s(n)$ is obtained at step $n+D$, as she has additional observations to draw on. Accordingly, the accuracy of this estimate is 
\begin{equation}   \eta(n)=\delta\left(s(n),\argmax_{s'\in\mc{S}}\left[\bm{\phi}_E(n+D;D)\right](s')\right).
\label{eq:eve_accuracy}
\end{equation}
This setup gives Eve an advantage, as she can wait up to $D$ steps before estimating the process state, while Bob's estimation is required to be performed in a timely fashion.

%We then need to distinguish between monitoring and control applications: in the former, $L_E(n;D)=0$ if $\phi_E(n;D)=\bm{\mu}$, i.e., Eve does not have any additional information about the Markov source than the steady-state distribution.
%On the other hand, the steady-state distribution in a remote control system depends on Bob's actions.
%As we will discuss in detail in Sec.~\ref{ssec:ade}, using a \gls{pp} can minimize leakage, as the timing of update requests is independent from the state of the system.
%In this case, we will use the entropy of the steady-state distribution under the optimal periodic policy as a normalization factor, i.e., we set $H_0=\bm{\mu}(\bm{\pi}^*_{\text{PP}})$.

\subsection{Forward-Backward State Estimation}

Since Eve sees the system as an \gls{hmm}, the \gls{map} estimate of the process state can be computed through the forward-backward algorithm.
Practically, Eve combines forward state-transition probabilities, which only consider the past, with backward state-transition probabilities, which only consider the future.
When estimating the state at time $m$ using information up to time $n>m$, the forward probabilities are based on observations from $0$ to $m$, while the backward probabilities are based on those from $m+1$ to $n$.

Upon observing the $k$-th request from Bob, Eve can recursively compute the forward probability for any possible initial state $s$ as
\begin{equation}f_k(s)=\sum_{s'\in\mc{S}}\zeta_{\tau(k),s}(s')\delta(\tau(k),\sigma(s'))f_{k-1}(s'),
\end{equation}
where $\tau(k)$ is the number of steps between transmissions $k-1$ and $k$, $\bm{\zeta}_{\Delta,s}$, given by~\eqref{eq:zeta}, is the state probability distribution in $\Delta$ steps assuming that the initial state was $s$, and $\sigma(\cdot)$ is the transmission scheduling policy defined in~\eqref{eq:timing_policy}. Recursion starts setting the initial probability vector $\mb{f}_0$ equal to the steady-state probability distribution, i.e., $\mb{f}_0=\bm{\mu}_0$. 

The backward probability for the same state is instead
\begin{equation}
    b_k(s;n)=\delta(\tau(k+1),\sigma(s))\sum_{s'\in\mc{S}}\zeta_{\tau(k+1),s}(s')b_{k+1}(s';n).
\end{equation}
The last step in the recursive calculation uses $b_{K(n)}(s)=|\mc{S}|^{-1}\,\forall$ $s\in\mc{S}$, as Eve has no information after index $K(n)$, which represents the index of the last transmission before time step $n$.
Eve's \gls{map} estimate of the process status when the $k$-th update is transmitted is then
\begin{equation}
    \phi_k(s;n)=\frac{f_k(s)b_k(s;n)}{\sum_{s'\in\mc{S}}f_k(s')b_k(s';n)}.
\end{equation}
%Eve can also estimate states in between transmission steps.
Eve can also compute the \gls{map} estimate of the process status $\ell$ steps after the $k$-th transmission step as
%and $\tau(k+1)-\ell$ steps before the $k+1$-th transmission step as
\begin{equation}
    \phi_{k}^{(\ell)}(s;n)=\sum_{\mathclap{s',s''\in\mc{S}}}\phi_k(s';n)\phi_{k+1}(s'';n)\zeta_{\ell,s'}(s)\zeta_{\tau(k+1)-\ell,s}(s'').
\end{equation}
Using the above formulas, Eve can compute the belief of the state distribution $\bm{\phi}_E(n;d)$ for any time step $n$ and delay $d$.
We observe that the running time of the forward-backward algorithm is $O(|\mc{S}|^2n)$. Therefore, it has a relatively low energy cost, which can be further reduced by limiting $n$ to the mixing time of the Markov chain.

\section{Eavesdropping Defenses}
\label{sec:defense}
While Bob aims to accurately estimate or control the process, limiting as much as possible the leakage of information, Eve is a purely adversarial attacker who wants to estimate the state of the remote Markov process exploiting the correlation between the state transitions of the system and the timing between Alice's transmissions.
%\subsection{Game Theoretical Framework}

%On the one hand, Eve aims to obtain information on the remote Markov process, taking advantage of the correlation between the state of the system and the timing between Alice's transmissions.
%On the other hand, Bob aims to accurately estimate or control the process while limiting as much as possible the leakage of information.
For a given opacity time gap $D$, the performance of the system can be defined as the expected weighted difference between the overall reward and the information leakage, i.e.,  
\begin{equation}
  \E{\sum_{n=0}^\infty R(s(n),c(n),a(n))-\varepsilon L_E(n;D)},
\label{eq:total_performance}
\end{equation}
where $\varepsilon>0$ is a parameter that can be used to adjust the relative importance of information leakage with respect to Bob's estimation accuracy.
Therefore, Bob's optimal strategy should maximize \eqref{eq:total_performance}, while Eve's best response consists of using the forward-backward algorithm to update her estimate of the Markov process.

We can model this system as a zero-sum \gls{oposg}~\cite{horak2023solving}.
The solution for the game is a \gls{ne} where any unilateral deviation from a player's policy would result in a decrease in that player's performance. Methods to solve zero-sum \glspl{oposg} have recently been proposed, based on the convexity property of the value function~\cite{horak2023solving} or on dividing the problem into sub-games with limited trajectories~\cite{delage2023hsvi}. However, complexity grows exponentially with the state space size. In fact, we can prove the following statement from well-known results in game theory. 
\begin{theorem}
    The computational time to find the \gls{ne} of the zero-sum game between Bob and Eve grows exponentially with the size $|\mc{S}|$ of the state space.
\end{theorem}
\begin{IEEEproof}
    A classical result by Dantzig~\cite{dantzig1951proof} proves that a two-player zero-sum game with payoff matrix $\mb{M}$ is equivalent to the following linear programming problem:
    \begin{equation}
        \begin{aligned}
            \text{minimize }&\sum_i \mb{x}\quad          \text{such that }\mb{x}\geq0,\ \mb{M}\mb{x}=1.
        \end{aligned}
    \end{equation}
    Normalizing $\mb{x}$ returns the optimal mixed strategy for one of the players. In our case, the action space for Bob is equivalent to the possible communication and control policies that he can adopt, which grows at least exponentially with the number of states $|\mc{S}|$. The length of $\mb{x}$ will then also grow exponentially with $|\mc{S}|$, making the game unsolvable in polynomial time.
\end{IEEEproof}
Although finding an \gls{ne} is computationally intractable for nontrivial problem sizes, we can design simple heuristic policies that allow Bob to trade-off between communication efficiency and system secrecy, reducing the vulnerability of \gls{goc} strategies to timing attacks.
In the following, we propose two solutions to attain this objective:
\acrfull{ade}, which alternates between goal-oriented and periodic transmission, and \acrfull{pde}, which is designed to reduce the entropy of Bob's scheduling decisions, thus increasing the communication opacity and making the system inherently more secure.

\subsection{Alternating Defense}\label{ssec:ade}

We know that the optimal \gls{goc} scheduling policy outperforms the optimal \gls{pp} in terms of expected reward, i.e., it can obtain the minimum transmission cost for a given state-estimation accuracy~\cite[Th. 2]{talli2024pragmatic}.
However, \gls{goc} is highly vulnerable to timing attacks, while a periodic strategy minimizes information leakage, as we prove below.
\begin{theorem}
    In an estimation scenario over a recurrent Markov chain, any periodic scheduling policy is perfectly private, i.e., the information leakage tends to the minimum value $L_{\text{min}}$ as $n$ increases for any finite value of $D$.
\end{theorem}
\begin{IEEEproof}
Under a periodic scheduling policy with period $T$, we have $\sigma(s)=T\ \forall\, s \in \mc{S}$ and, consequently, the forward probabilities are $f_k(s)=\sum_{s'\in\mc{S}}\left(\mb{P}^T\right)_{s',s}f_{k-1}(s')$.
This is exactly equivalent to a blind update, and the same holds for the backward probabilities.
As timing does not provide new information, Eve's belief tends to the steady-state distribution $\bm{\mu}$ for any $n$ larger than the system mixing time, reducing the leakage to $L_{\text{min}}$, defined in \eqref{eq:leak_min}, as the window for the leakage calculation moves past the initial transient.
\end{IEEEproof}
We note that the theorem may not always hold in the more general control case, as Bob's control policy $\pi$ affects the steady-state distribution $\bm{\mu}$.
However, the general principle holds, as periodic transmission strategies still minimize leakage for any sequence of control decisions.

% \begin{figure}[t]
% \begin{algorithm}[H]
% \caption{\acrfull{ade}}
% \label{alg:ade}
% \begin{algorithmic}[1]
% \footnotesize

% \Function{Schedule}{$s, \sigma,T,\mb{P}, \mb{f}, \mb{b}, \bm{\tau}, L_{\text{low}}, L_{\text{high}},\xi$}
% \State $L_{\text{goc}}\gets L_E$ with $\tau(k)=\sigma(s)$
% \State $L_{\text{per}}\gets L_E$ with $\tau(k)=T$
% \If{$\xi=0$} \Comment{Goal-oriented scheduling active}
%     \If {$L_{\text{goc}}\geq L_{\text{high}}$} \Comment{Check privacy threshold}
%     \State \Return{$T,1$}\Comment{Switch to PP}
%     \Else
%     \State \Return{$\sigma(s),0$}\Comment{Keep using MPI}
%     \EndIf
% \Else \Comment{Periodic scheduling active}
%     \If {$L_{\text{per}}< L_{\text{low}}$} \Comment{Check performance threshold}
%     \State \Return{$\sigma(s),0$}\Comment{Switch to MPI}
%     \Else
%     \State \Return{$T,1$}\Comment{Keep using PP}
%     \EndIf
% \EndIf
% \EndFunction
% \end{algorithmic}
% \end{algorithm}
% \end{figure}

\begin{figure}[t]
\vspace{-8pt}
\begin{algorithm}[H]
\caption{\acrfull{ade}}
\label{alg:ade}
\begin{algorithmic}[1]
\footnotesize

\Function{Schedule}{$s, \sigma,T,\mb{P}, \mb{f}, \mb{b}, \bm{\tau}, L_{\text{low}}, L_{\text{high}},\xi$}
\If{$\xi=0$} \Comment{\gls{goc}  active}
    \If {$L_{E}(\sigma(s))\geq L_{\text{high}}$} \Comment{Check secrecy threshold}
    \State \Return{next update in $T$ steps, $\xi=1$}\Comment{Switch to \gls{pp}}
    \Else
    \State \Return{next update in $\sigma(s)$ steps, $\xi=0$}\Comment{Keep using \gls{goc}}
    \EndIf
\Else \Comment{\gls{pp} active}
    \If {$L_{E}(T)< L_{\text{low}}$} \Comment{Check performance threshold}
        \State \Return{next update in $\sigma(s)$ steps, $\xi=0$}\Comment{Switch to \gls{goc}}
\Else
        \State \Return{next update in $T$ steps, $\xi=1$}\Comment{Keep using \gls{pp}}
    \EndIf
\EndIf
\EndFunction
\end{algorithmic}
\end{algorithm}
\vspace{-0.3cm}
\end{figure}

We take advantage of this principle to design our first heuristic policy, \acrfull{ade}, whose pseudocode is reported as Algorithm~\ref{alg:ade}.
As Bob knows his own transmission policy and, hence, the timing signal observed by Eve, he can compute the information leakage during the next transmission interval.
Hence, Bob can switch to a \acrfull{pp} whenever the expectation of future leakage increases beyond an upper threshold $L_{\text{high}}$ and switch back to \gls{goc} whenever the future leakage goes below a threshold $L_{\text{low}}$.
This hysteresis pattern allows Bob to limit both the average and maximum leakage, while still exploiting \gls{goc} at least in some time intervals.

\subsection{Packing Defense}

The second heuristic policy is named \acrfull{pde} and is based on a simple observation: if multiple states are mapped to the same inter-transmission period, the leakage of the timing signal decreases, as Eve has a harder time distinguishing between states.
We then define the entropy of the scheduling policy $\sigma$ as
\begin{equation}\label{eq:policy_entropy}
    H(\sigma)=-\sum_{\tau=1}^{\infty}\frac{\sum_{s\in\mc{S}} \delta(\tau,\sigma(s))}{|\mc{S}|}\log_2\left(\frac{\sum_{s\in\mc{S}} \delta(\tau,\sigma(s))}{|\mc{S}|}\right).
\end{equation}
We can assume that $H(\sigma)$ is a good proxy for leakage: any periodic policy has zero entropy, while the maximum entropy $\log_2(|\mc{S}|)$ is achieved by picking a different inter-transmission interval for each state, i.e., when $\forall$ $s', s'' \in \mc{S}$, $s' \neq s''$, we have $\sigma(s') \neq \sigma(s'')$.
In this case, any timing signal $\tau$ is mapped to a different state, so that at each transmission Eve gains perfect knowledge of the transmitted value. 

To define the \gls{pde} strategy, we introduce the concept of \textit{single-state deviation policy} $\xi_{\sigma}^{(s^*,\tau)}$, which is a scheduling strategy identical to $\sigma$ except for state $s^*$, whose associated scheduling period is set to $\tau$: 
\begin{equation}
    \xi_{\sigma}^{(s^*,\tau)}(s)=\tau\delta(s,s^*)+\sigma(s)(1-\delta(s,s^*)).
\end{equation}
Starting from the purely goal-oriented policy, denoted by $\sigma^{(0)}$, we can then define an iterative procedure to \emph{pack} the policy through a series of single-state deviations that gradually reduce the entropy. The $i$-th packing iteration is defined as $\sigma^{(i)}(s)=   \xi_{\sigma^{(i-1)}}^{(s_{i}^*,\tau_i)}(s)$ for all $s\in S$, where
\begin{equation}
(s^*_i,\tau_i)=\argmax_{(s^*,\tau):H\left(\xi_{\sigma^{(i-1)}}^{(s^*,\tau)}\right)<H(\sigma^{(i-1)})}\E{R_B|\xi_{\sigma^{(i-1)}}^{(s^*,\tau)}}\,.
\end{equation}
% \begin{equation}
%     \sigma^{(i+1)}=\argmax_{s^*\in\mc{S},\tau\in\mathbb{N}:H\left(\xi_{\sigma^{(i)}}^{(s^*,\tau)}\right)<H(\sigma^{(i)})}\E{R_B|\xi_{\sigma^{(i)}}^{(s^*,\tau)}},
% \end{equation}
This packing rule ensures that the new policy $\sigma^{(i)}$ is the one that maximizes the expected system reward $\E{R_B}$ among those with entropy lower than $H\left(\sigma^{(i-1)}\right)$. 
We can repeat the packing step until the final policy achieves a target entropy value $H^*$, which represents the stopping criterion for \gls{pde}.
The full \gls{pde} pseudocode is given in Algorithm~\ref{alg:pde}.

\begin{figure}[t]
\vspace{-8pt}
\begin{algorithm}[H]
\caption{\acrfull{pde}}
\begin{algorithmic}[1]
\footnotesize

\Function{Pack}{$\sigma, H^*$}
\State $H\gets$\Call{entropy}{$\sigma$} \Comment{Compute entropy using~\eqref{eq:policy_entropy}}
\State running $\gets$ true
\While{running}
  \State running $\gets$ false
  \State $R\gets -\infty$
  \State $\sigma'\gets\sigma$
  \For{$s^*\in\mc{S}$}
    \For{$\tau\in\{1,\ldots,T_{\text{max}}\}$}
      \If{\Call{entropy}{$\xi_{\sigma}^{(s^*,\tau)}$}$<H$}\Comment{Check entropy}
        \If{\Call{reward}{$\xi_{\sigma}^{(s^*,\tau)}$}$>R$}
          \State $\sigma'\gets\xi_{\sigma}^{(s^*,\tau)}$
          \State $R\gets$\Call{reward}{$\sigma'$}
        \EndIf
      \EndIf
    \EndFor
  \EndFor
  \If{$\sigma'\neq\sigma$}
    \State $\sigma\gets\sigma'$\Comment{Update policy}
    \State $H\gets$\Call{entropy}{$\sigma$}
    \If {$H>H^*$} \Comment{Stopping criterion}
      \State running $\gets$ true
    \EndIf
  \EndIf
\EndWhile
\EndFunction
\end{algorithmic}
\label{alg:pde}
\end{algorithm}
\vspace{-0.3cm}
\end{figure}

\section{Simulation Settings and Results}
\label{sec:results}

In the following, we study our \gls{goc} model in two simulation scenarios.
The first represents a \textit{remote estimation} task, where Bob aims to estimate the current state of the system, which evolves independently from Bob's actions.
The second is a \textit{remote control} task in which Bob affects the evolution of the system with the goal of reaching certain states. 
After presenting each scenario, we analyze the performance of the heuristic policies introduced in Sec.~\ref{sec:defense} against the optimal \gls{goc} scheduling, computed via the \acrfull{mpi} algorithm, and the optimal \acrfull{pp}. 

\subsection{Scenario Settings}

The remote estimation and remote control scenarios are both modeled according to the discrete time \gls{pomdp} presented in Sec.~\ref{sec:model}.
Although the proposed framework is valid for any recurrent Markov chain, we focus on a class of processes that allow for an easy analysis of the system's behavior under different conditions.
We consider a state space of $|\mc{S}|=30$ states, numbered from $1$ to $30$.
The transition probability function $P: \mc{S} \times \mc{S} \times \mc{A} \rightarrow [0,1]$ (corresponding to the matrix $\bm{P}$) depends on a single parameter $\theta$ named \textit{density decay}, that makes it possible to tune the predictability of the evolution of the system.
Specifically, we have
\begin{equation}
P (s, s^\prime, a) = 
    \begin{cases}
       \frac{2 - 2 g(s, \theta)}{6}, & s^\prime = \chi(s,a) \oplus 1 ,\, \Mod(s,4)= 2; \\
        \frac{2 + g(s, \theta)}{6}, & s' = \chi(s,a) \oplus3,\, \Mod(s,4)= 2;\\
        \frac{2 + g(s, \theta)}{6}, & s^\prime = \chi(s,a) \ominus2,\, \Mod(s,4)= 2;\\
        \frac{1 + 2 g(s, \theta)}{3}, & s^\prime = \chi(s,a) \oplus 1 ,\, \Mod(s,4)\neq 2; \\
        \frac{1 - g(s, \theta)}{3}, & s^\prime = \chi(s,a) \oplus3,\, \Mod(s,4)\neq 2;\\
        \frac{1 - g(s, \theta)}{3}, & s^\prime = \chi(s,a) \ominus2,\, \Mod(s,4)\neq 2;\\
        0, & \text{otherwise};
    \end{cases}
    \label{eq:transition_prob}
\end{equation}
where $\oplus$ and $\ominus$ represent modulo $|\mc{S}|$ addition and subtraction, $\Mod(m,n)$ is the integer modulo function, $\theta \in \mathbb{R}^+$ is the density decay, and $g(s, \theta)$ is defined as
\begin{equation}
    g(s, \theta) =\left| \frac{2(s-2)}{|\mathcal{S}|-2}-1  \, \right|^{\theta}\in[0,1].
    \label{eq:transition_std}
\end{equation}
% \begin{equation}
%     g(s, \theta) =\left(| 2i - |\mathcal{S}| \cdot |\right)^{\theta}|\mathcal{S}|^{-\theta}.
%     \label{eq:transition_std}
% \end{equation}
The function $\chi(s,a) \in \mc{S}$ determines the state transition associated with action $a \in \mc{A}$, which is $\chi(s,a)=s$ in remote estimation (therefore, independent of Bob's actions), and$\chi(s,a)=s+a$ in the case of remote control. 

Hence, from any state $s$, transitions can occur with a non-zero probability to only three landing states that, only for the control scenario, depend on the action $a$. 
The probabilities of moving to the farthest reachable states ($\chi(s,a)\oplus 3$ or $\chi(s,a)\ominus 2$) are always balanced.
Instead, the transition to the intermediate state $\chi(s,a)\oplus 1$ is more probable than the other two transitions from all states, except those such that $\Mod(s,4)=2$, making the drift of the process more variable.
We observe that as $\theta \rightarrow \infty$, $g(s, \theta)$ tends to zero, and the transition probabilities to neighboring states will become more uniform (and less predictable).
Conversely, as $\theta \rightarrow 0$, $g(s, \theta)$ tends to $1$ and most states will have deterministic (and, hence, fully predictable) transitions. 
Finally, we note that $g(s, \theta)=1$ for the extreme states $s=1$ and $s=|S|$, and progressively decreases when moving towards the middle states. For any value of $\theta$, middle states tend to have more balanced transition probabilities toward their landing states, while states closer to the extremes have more unbalanced transition probabilities, that is, more predictable transitions.

As already mentioned, 
%Bob's action in the estimation scenario consists of estimating the current state of the remote process.
%Therefore, 
Bob's action space $\mc{A}$ in the estimation scenario is identical to the state space, and the task reward function is $r_B(s, a)=\delta(s, a)$. 
In the remote control scenario, the action space is $\mc{A}=\{0, 1, 2\}$ and we defined $\chi(s,a)=s+a$. Therefore, Bob can (stochastically) control the sequence of states by choosing proper actions. 
% landing states associated with each state $s \in \mc{S}$ by taking .
In our experiments, we assumed the control goal was to keep the remote process  close to the middle state $s^\circ=14$. Accordingly, we define the reward as $r_B(s,a)= 5 \cdot \exp( - \left| s - s^\circ \right|)$, $\forall$~$s \in \mathcal{S}$.
Note that the control reward does not depend on the accuracy of Bob's estimates but only on the distance between the current state $s$ and the target state $s^\circ$.
%Specifically, we set $s^\circ=14$, so that Bob's goal is to keep the process in the middle state. 

In both scenarios, we generate multiple \gls{pomdp} configurations, varying the density decay $\theta \in [1, 2^7]$ and the transmission cost $\beta \in [0.2, 2]$. 
For each configuration, we compute the optimal \gls{goc} scheduling policy given by the \gls{mpi} algorithm~\cite{talli2024pragmatic}, maximizing the long-term reward of the system penalized by the communication cost, as defined in \eqref{eq:return_effective}. 
Hence, we compare \gls{mpi} with \gls{pp}, which is the best policy among those exploiting a fixed inter-transmission period, and the two heuristics presented in Sec.~\ref{sec:defense}. 
The \gls{pde} heuristic is configured by setting $H^*=\frac{1}{2} H\left(\sigma^{(0)}\right)$ as a stopping criterion, where $\sigma^{(0)}$ is the initial scheduling policy returned by the \gls{mpi} algorithm. 
Instead, the \gls{ade} heuristic uses $L_{\text{low}}=0.4$ and $L_{\text{high}}=0.6$ as leakage thresholds. 
In all cases, we set $T_{\text{max}} = 10$ as the maximum interval between consecutive transmissions, i.e., the maximum value that the scheduling function $\sigma(\cdot)$ can take.

\subsection{Remote Estimation Scenario}

\begin{figure}[t!]
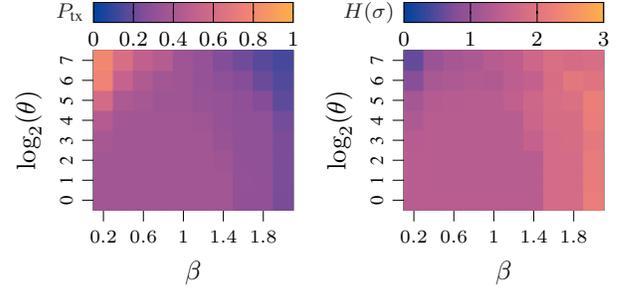

    \centering
    \subfloat[Transmission probability.~\label{fig:mpi_transmission_estimation}]
    {\input{./tikz_fig/estimate/transmission_effective.tex}}
    \subfloat[Scheduling policy entropy.~\label{fig:mpi_entropy_estimation}]
    {\input{./tikz_fig/estimate/entropy_effective.tex}}
    \caption{Characterization of the \gls{mpi} policy as a function of $\beta$ and the density decay $\theta$ in the estimation scenario.}
    \label{fig:mpi_policy_estimation}
\end{figure}

Focusing on the remote estimation scenario, we first analyze the characteristics of the optimal \gls{goc} policy provided by the \gls{mpi} algorithm. 
Fig.~\ref{fig:mpi_transmission_estimation} shows a heatmap of the transmission probability associated with each system configuration.
We can see that transmissions become less likely as $\beta$ increases and are also affected by the randomness of the system's evolution, which depends on the density decay $\theta$.
When the transmission cost is low ($\beta\to 0$), larger values of $\theta$ (which correspond to less predictable transition matrices) result in more frequent state update requests from Bob, who can exploit communication to keep track of the process evolution. However, if the transmission cost increases ($\beta \rightarrow 2$), the trend reverts and the transmission probability decreases as $\theta$ increases, because the higher estimation accuracy may not cover the update cost. 

Fig.~\ref{fig:mpi_entropy_estimation} represents the entropy $H(\sigma)$ of the transmission scheduling policy returned by the \gls{mpi} algorithm, which is a proxy of information leakage caused by transmission decisions. 
In general, $H(\sigma)$ decreases as $\beta \rightarrow 0$, because when the frequency of communication increases, the variability of the inter-transmission time decreases, and Eve has more difficulty in sorting out the states sequence from the timing signal.  
This phenomenon is more evident for $\theta \rightarrow 2^7$, which represents a condition in which state transitions are less predictable and the optimal scheduling becomes similar to the \gls{pp} strategy.

\begin{figure}[t!]
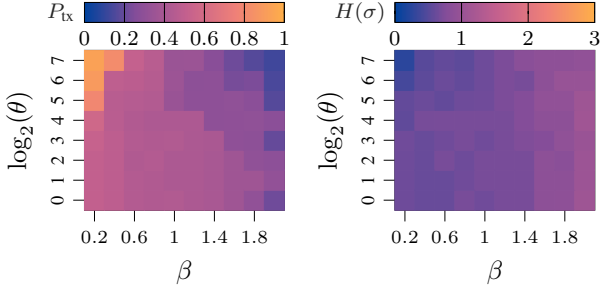

    \centering
    \subfloat[Transmission probability.~\label{fig:pde_transmission_estimation}]
    {\input{./tikz_fig/estimate/transmission_pde.tex}}
    \subfloat[Scheduling policy entropy.~\label{fig:pde_entropy_estimation}]
    {\input{./tikz_fig/estimate/entropy_pde.tex}}
    \caption{Characterization of the \gls{pde} policy as a function of $\beta$ and the density decay $\theta$ in the estimation scenario.}
    \label{fig:pde_policy_estimation}
\end{figure}

In general, a policy that selects a different value of $\sigma(s)$ for each state would have an entropy equal to $\log_2(|\mc{S}|)$, while any periodic policy would have zero entropy.
Hence, we expect the \gls{mpi} algorithm to have the highest entropy $H(\sigma)$ among the strategies analyzed in this paper.
This is because \gls{pp} uses a fixed inter-transmission period, \gls{ade} alternates between \gls{mpi} and \gls{pp}, while \gls{pde} is explicitly designed to reduce $H(\sigma)$ with respect to \gls{mpi}.
We can appreciate the advantages of \gls{pde} by looking at Fig.~\ref{fig:pde_policy_estimation}, where we report the transmission probability and entropy associated with the heuristic. 
Notably, \gls{pde} results in a transmission probability similar to \gls{mpi} but successfully halves the entropy in all configurations of the system. 

\begin{figure}[t!]
    \centering
    \input{./tikz_fig/estimate/leakage_time.tex}
    \caption{Information leakage during a single episode in the estimation scenario, with $\beta=1$, $\theta=32$ and $D=5$. The \gls{ade} thresholds $L_{\text{low}}$ and $L_{\text{high}}$ are marked as dashed lines.}
 \label{fig:leak_vs_time_estimation}
\end{figure}

\begin{figure}[t!]
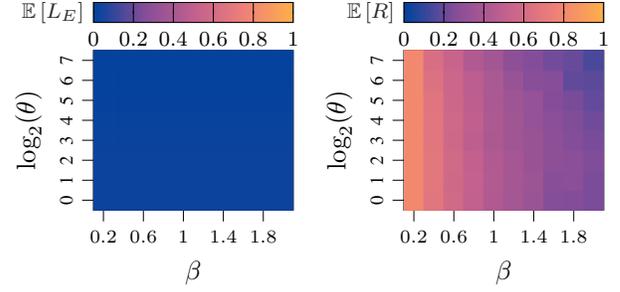
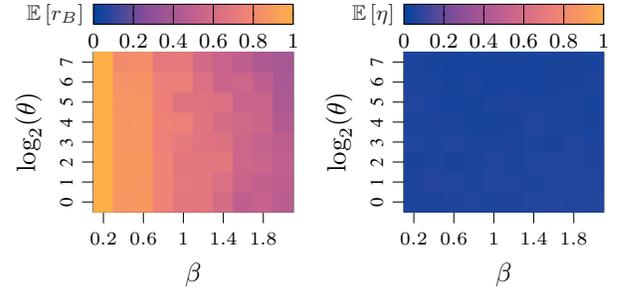

    \centering
    \subfloat[Leakage.~\label{fig:leak_per_estimation}]
    {\input{./tikz_fig/estimate/leakage_periodic.tex}}
    \subfloat[Total reward.~\label{fig:rew_per_estimation}]
    {\input{./tikz_fig/estimate/reward_periodic.tex}}\\
    \subfloat[Estimation reward.~\label{fig:bob_per_estimation}]
    {\input{./tikz_fig/estimate/acc_bob_periodic.tex}}
    \subfloat[Eve's accuracy.~\label{fig:eve_per_estimation}]
    {\input{./tikz_fig/estimate/acc_eve_periodic.tex}}    
    \caption{\gls{pp} performance as a function of $\theta$ and $\beta$ in the estimation scenario, with $D=5$.}
    \label{fig:pp_heatmaps_estimation}
\end{figure}

Although $H(\sigma)$ is a useful indicator of system opacity, the information leakage $L_E$, as defined in \eqref{eq:leakage}, provides more information on the trade-off between secrecy and performance of the remote estimation task.
In Fig.~\ref{fig:leak_vs_time_estimation} we then report  $L_E$ during a single episode of $N_{\text{step}}=200$ steps, considering $\beta=1$, $\theta=32$, and $D=5$.
In addition to \gls{mpi} and \gls{pde}, we also consider the optimal periodic strategy \gls{pp} and the \gls{ade} heuristic.
We can observe that the leakage of the \gls{mpi} algorithm quickly approaches $1$, showing that Eve correctly guesses the remote state very often with these settings.
Conversely, \gls{pp} does not provide any information to Eve, whose knowledge is limited to the steady-state probability distribution of the Markov process.
By design, the \gls{ade} algorithm keeps the leakage  between $L_{\text{low}}$ and $L_{\text{high}}$, thus offering a compromise between the two previous approaches.
Finally, \gls{pde} improves secrecy compared to \gls{mpi}, but the value of $L_E$ exhibits strong oscillations over time, exposing the system to a high risk of leakage in some steps.

\begin{figure}[t!]
    \subfloat[Leakage.~\label{fig:leak_eff_estimation}]
    {\input{./tikz_fig/estimate/leakage_effective.tex}}
    \subfloat[Total reward.~\label{fig:rew_eff_estimation}]
    {\input{./tikz_fig/estimate/reward_effective.tex}}\\
    \subfloat[Estimation reward.~\label{fig:bob_eff_estimation}]
    {\input{./tikz_fig/estimate/acc_bob_effective.tex}}
    \subfloat[Eve's accuracy.~\label{fig:eve_eff_estimation}]
    {\input{./tikz_fig/estimate/acc_eve_effective.tex}}
    \caption{\gls{mpi} performance as a function of $\theta$ and $\beta$ in the estimation scenario, with $D=5$.}
    \label{fig:mpi_heatmaps_estimation}
\end{figure}

Fig.~\ref{fig:pp_heatmaps_estimation} shows the performance of \gls{pp} while varying the communication cost $\beta$ and density decay $\theta$, and considering a total of $N_{\text{ep}}=10$ episodes for each configuration, with $N_{\text{step}}=200$.
In addition to leakage (a), we consider the total reward $R$ (b), defined in \eqref{eq:effective_reward}, the reward for the estimation task $r_B$ (c), as well as the probability $\eta$ that Eve correctly estimates the state of the Markov process (d), the latter defined as in \eqref{eq:eve_accuracy}.
First, we observe that $L_{E} \approx 0$ for all system configurations when the \gls{pp} solution is used, as expected for periodic communication. 
% as we discussed in Sec.~\ref{sec:defense}, periodic communication is fully opaque to timing attacks.
%Moreover, Fig.~\ref{fig:bob_per_estimation} shows that $r_B$, which represents Bob's accuracy in estimating the Markov source, decreases as transmissions become less frequent. 
%Finally, the expected total reward, shown in Fig.~\ref{fig:rew_per_estimation}, degrades in the case of high communication cost ($\beta \rightarrow 2$) and stochastic transitions ($\theta \gg 1$).
Moreover, from Fig.~\ref{fig:rew_per_estimation} and Fig.~\ref{fig:bob_per_estimation} we observe that the expected total reward $\E{R}$, as well as Bob's state estimation accuracy $r_B$, decrease for larger transmission costs ($\beta$, which yield longer inter-transmission periods) and more erratic transition probabilities ($\theta \gg 1$).

Fig.~\ref{fig:mpi_heatmaps_estimation} offers a comparison of these performance indicators for the \gls{mpi} strategy that, being purely \gls{goc}, is complementary to \gls{pp}. Not surprisingly, this setting leads to a strong secrecy degradation (Fig.~\ref{fig:leak_eff_estimation}): the information leakage is close to $0.8$ for all configurations except for those with very high values of $\beta$ and $\theta$.
Eve is able to correctly decode the status of the monitored process almost as often as Bob (Fig.~\ref{fig:eve_eff_estimation}), highlighting the strong vulnerability of \gls{mpi} to timing attacks.
On the other hand, \gls{mpi} significantly improves the total reward compared to \gls{pp} (Fig.~\ref{fig:rew_eff_estimation}).
Since \gls{mpi} tends to transmit more often than \gls{pp}, the accuracy of Bob does not decrease significantly as $\beta$ increases and the gain over \gls{pp} reaches $50\%$ when $\beta \rightarrow 2$.

\begin{figure}[t!]
    \subfloat[Leakage.~\label{fig:leak_heu_estimation}]
    {\input{./tikz_fig/estimate/leakage_heuristic.tex}}
    \subfloat[Total reward.~\label{fig:rew_heu_estimation}]
    {\input{./tikz_fig/estimate/reward_heuristic.tex}}\\ 
    \subfloat[Estimation reward.~\label{fig:bob_heu_estimation}]
    {\input{./tikz_fig/estimate/acc_bob_heuristic.tex}}
    \subfloat[Eve's accuracy.~\label{fig:eve_heu_estimation}]
    {\input{./tikz_fig/estimate/acc_eve_heuristic.tex}}
    \caption{\gls{ade} performance as a function of $\theta$ and $\beta$ in the estimation scenario, with $D=5$.}
    \label{fig:ade_heatmaps_estimation}
\end{figure}

\begin{figure}[t!]
    \subfloat[Leakage.~\label{fig:leak_pde_estimation}]
    {\input{./tikz_fig/estimate/leakage_pde.tex}}
    \subfloat[Total reward.~\label{fig:rew_pde_estimation}]
    {\input{./tikz_fig/estimate/reward_pde.tex}}\\ 
    \subfloat[Estimation reward.~\label{fig:bob_pde_estimation}]
    {\input{./tikz_fig/estimate/acc_bob_pde.tex}}
    \subfloat[Eve's accuracy.~\label{fig:eve_pde_estimation}]
    {\input{./tikz_fig/estimate/acc_eve_pde.tex}}
    \caption{\gls{pde} performance as a function of $\theta$ and $\beta$ in the estimation scenario, with $D=5$.}
    \label{fig:pde_heatmaps_estimation}
\end{figure}

The proposed heuristic strategies are expected to perform somewhere between \gls{pp} and \gls{mpi}.
As shown in Fig.~\ref{fig:leak_heu_estimation}, \gls{ade} improves secrecy in all configurations, guaranteeing that the leakage remains lower than  $L_{\text{high}}$.
Comparing Fig.~\ref{fig:bob_heu_estimation} and Fig.~\ref{fig:eve_heu_estimation}, we note that Eve's accuracy is much lower than Bob's, unlike in the \gls{mpi} scenario, leading to a mean leakage of $0.45$.
At the same time, Fig.~\ref{fig:rew_heu_estimation} shows that \gls{ade} degrades the total reward compared to \gls{mpi}, especially in the case of Markov chains with low $\theta$ and high transmission cost.
On the other hand, the reward of \gls{ade} presents a performance gain of approximately $10\%$ over \gls{pp}, as apparent from the comparison between Fig.~\ref{fig:rew_heu_estimation} and Fig.~\ref{fig:rew_per_estimation}.

In Fig.~\ref{fig:pde_heatmaps_estimation}, we report the results of \gls{pde}, which, similarly to \gls{ade}, strikes a compromise between the higher efficiency of \gls{mpi} and the secrecy provided by periodic scheduling. 
The main difference is that \gls{pde} does not monitor information leakage explicitly, but considers the entropy of the scheduling policy as a secrecy indicator. 
Fig.~\ref{fig:leak_pde_estimation} shows that the expected leakage with \gls{pde} is higher than with \gls{ade} (Fig.~\ref{fig:eve_pde_estimation}) without significantly improving Bob's performance.
Hence, \gls{ade} performs better than \gls{pde} in this remote estimation task; however, the higher computational complexity may make \gls{ade} unsuitable for implementation on nodes with limited hardware.

\begin{figure}[t!]
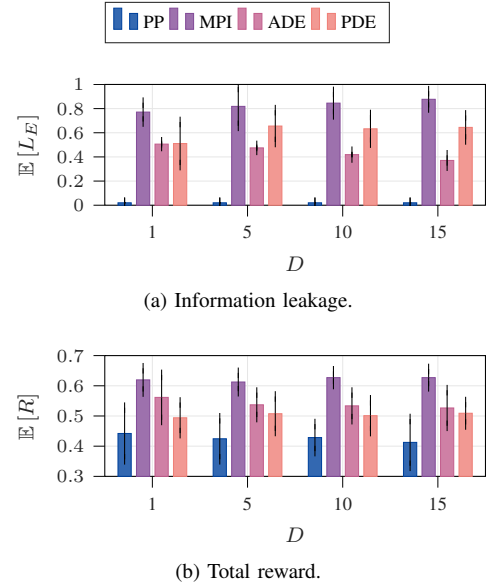

    \centering
    \subfloat{\input{tikz_fig/estimate/delay_legend}}\\
    \setcounter{subfigure}{0}
    \subfloat[Information leakage.~\label{fig:delay_leak}]
    {\input{./tikz_fig/estimate/delay_leakage.tex}}\\
    \subfloat[Total reward.~\label{fig:delay_rew}]
    {\input{./tikz_fig/estimate/delay_reward.tex}}
    \caption{Expected leakage and reward as a function of $D$ in the estimation scenario, with $\beta=1$ and $\theta=32$.}
    \label{fig:delay_impact_estimation}
\end{figure}

Fig.~\ref{fig:delay_impact_estimation} analyzes the impact of the time gap $D$ on overall performance, focusing on a system with $\beta=1$ and $\theta=32$, and setting $D\in\{1, 5, 10, 15\}$.
As expected, the performance of \gls{pp} does not change in the different scenarios, while the leakage of \gls{mpi} increases as a function of $D$, given that a longer time interval allows Eve to exploit more information.
As the \gls{pde} scheduling policy is directly derived from \gls{mpi}, \gls{pde} follows a similar trend in terms of leakage.
Instead, \gls{ade} tends to make more conservative choices and switches to \gls{pp} more often and for longer periods, as $D$ increases.
In particular, for $D=15$, the reward of \gls{ade} approximates that of \gls{pde} that, for all the other configurations, shows a worse performance. 

\subsection{Remote Control Scenario}

\begin{figure}[t!]
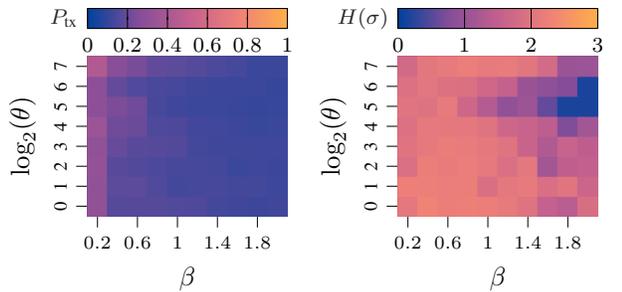

    \centering
    \subfloat[Transmission probability.~\label{fig:mpi_transmission_control}]
    {\input{./tikz_fig/control/transmission_effective.tex}}
    \subfloat[Entropy of the scheduling policy.~\label{fig:mpi_entropy_control}]
    {\input{./tikz_fig/control/entropy_effective.tex}}
    \caption{Characterization of the \gls{mpi} policy as a function of $\theta$ and $\beta$ in the control scenario.}
    \label{fig:mpi_policy_control}
\end{figure}

\begin{figure}[t!]
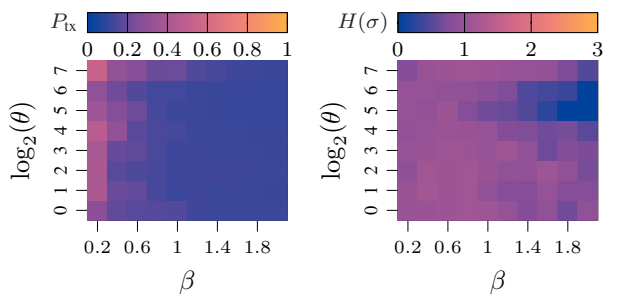

    \centering
    \subfloat[Transmission probability.~\label{fig:pde_transmission_control}]
    {\input{./tikz_fig/control/transmission_pde.tex}}
    \subfloat[Entropy of the scheduling policy.~\label{fig:pde_entropy_control}]
    {\input{./tikz_fig/control/entropy_pde.tex}}
    \caption{Characterization of the \gls{pde} policy as a function of $\theta$ and $\beta$ in the control scenario.}
    \label{fig:pde_policy_control}
\end{figure}

The remote control scenario has a significant difference with respect to the estimation scenario: Bob does not need to know the status of the process to maximize the reward, which depends on the closeness between the current state $s(n)$ and the target state $s^{\circ}$. 
This strongly reduces the transmission probability of the \gls{mpi} strategy compared to the estimation scenario.
As shown in Fig.~\ref{fig:mpi_transmission_control}, Bob updates his state estimate with high frequency only when the evolution of the process becomes highly stochastic ($\theta \gg 1$) or if the transmission cost is negligible ($\beta \rightarrow 0$).
A similar trend can be observed for the \gls{pde} policy, whose transmission rate is reported in Fig.~\ref{fig:pde_transmission_control} and results slightly higher than that of \gls{mpi}.

\begin{figure}[t!]
    \centering
    \input{./tikz_fig/control/leakage_time.tex}
    \caption{Information leakage during a single episode in the control scenario, with $\beta=1$, $\theta=32$ and $D=5$. The \gls{ade} thresholds $L_{\text{low}}$ and $L_{\text{high}}$ are marked as dashed lines.}
 \label{fig:leak_vs_time_control}
\end{figure}

\begin{figure}[t!]
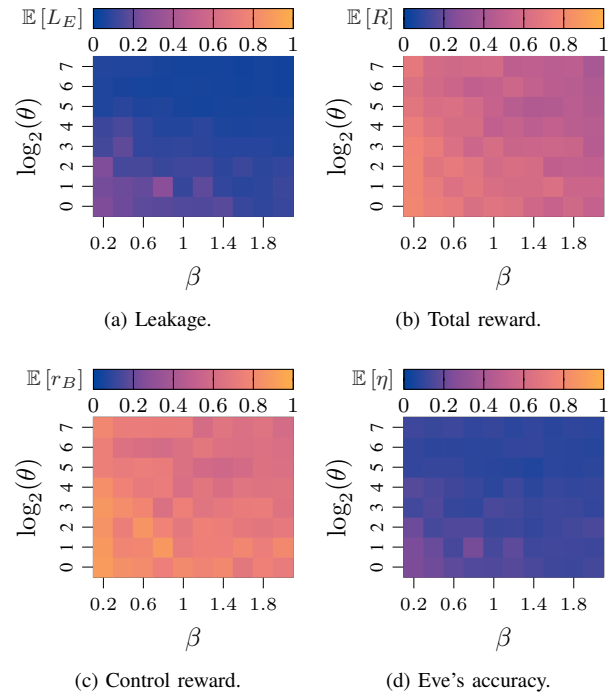

    \centering
    \subfloat[Leakage.~\label{fig:leak_per_control}]
    {\input{./tikz_fig/control/leakage_periodic.tex}}
    \subfloat[Total reward.~\label{fig:rew_per_control}]
    {\input{./tikz_fig/control/reward_periodic.tex}}\\
    \subfloat[Control reward.~\label{fig:bob_per_control}]
    {\input{./tikz_fig/control/acc_bob_periodic.tex}}
    \subfloat[Eve's accuracy.~\label{fig:eve_per_control}]
    {\input{./tikz_fig/control/acc_eve_periodic.tex}}    
    \caption{\gls{pp} performance as a function of $\theta$ and $\beta$ in the control scenario, with $D=5$.}
    \label{fig:pp_heatmaps_control}
\end{figure}

Fig.~\ref{fig:mpi_entropy_control} reports the entropy of the \gls{mpi} scheduling, which strongly decreases in configurations with $\theta > 16$ and $\beta > 1.4$.
We can hypothesize that, in such cases, requesting state updates from Alice is inconvenient and \gls{mpi} associates most states with the maximum inter-transmission interval $T_{\text{max}}$.
On the other hand, the entropy increases again for $\theta > 64$, denoting that the relation between the stochasticity of the system and the optimal scheduling decisions is more complex. 
Looking at Fig.~\ref{fig:pde_entropy_control}, we can appreciate how \gls{pde} follows the same pattern and, as we set $H^*=\frac{1}{2} H\left(\sigma^{(0)}\right)$, reduces the entropy of the \gls{mpi} scheduling policy by $50\%$.

As in the estimation case, we first focus on a single episode (with $\theta=32$, $\beta=1$, and $D=5$) and compare the leakage obtained by \gls{mpi}, \gls{pp}, and the two heuristic strategies. 
First, we observe that the leakage of \gls{pp} is constant but has higher values than in the estimation task.
This is because the steady-state distribution $\bm{\mu}(\pi)$ of the system presents higher entropy, as it is directly influenced by Bob's actions.
Indeed, Bob aims to keep the current state as close as possible to $s^\circ$, reducing the system's randomness, and consequently, increasing the leakage.
%independently of the scheduling decision.
We also note that the leakage of \gls{mpi} does not increase beyond $0.7$ and, consequently, \gls{ade} rarely switches to periodic communication, while \gls{pde} substantially improves secrecy with respect to both \gls{mpi} and \gls{ade}.

\begin{figure}[t!]
    \subfloat[Leakage.~\label{fig:leak_eff_control}]
    {\input{./tikz_fig/control/leakage_effective.tex}}
    \subfloat[Total reward.~\label{fig:rew_eff_control}]
    {\input{./tikz_fig/control/reward_effective.tex}}\\
    \subfloat[Control reward.~\label{fig:bob_eff_control}]
    {\input{./tikz_fig/control/acc_bob_effective.tex}}
    \subfloat[Eve's accuracy.~\label{fig:eve_eff_control}]
    {\input{./tikz_fig/control/acc_eve_effective.tex}}
    \caption{\gls{mpi} performance as a function of $\theta$ and $\beta$ in the control scenario, with $D=5$.}
    \label{fig:mpi_heatmaps_control}
\end{figure}

The fact that \gls{pp} may have a non-zero leakage in control tasks is confirmed by Fig.~\ref{fig:leak_per_control}, which reports the expected leakage $\E{L_E}$ for all combinations of density decay $\theta$ and communication cost $\beta$. 
Interestingly, the system is more vulnerable to timing attacks for $\beta \rightarrow 0$ and $\theta \rightarrow 0$, representing the case in which Markov transitions are more deterministic.  
The same configuration leads to an increase in the average reward $\E{r_B}$ of the control task and a slight increase of Eve's  accuracy $\E{\eta}$ (Fig.~\ref{fig:bob_per_control}-\subref*{fig:eve_per_control}). 

Fig.~\ref{fig:mpi_heatmaps_control} reports the same analysis for the \gls{mpi} approach.
Comparing Fig.~\ref{fig:leak_eff_control} and Fig.~\ref{fig:mpi_entropy_control}, we observe that the information leakage strongly decreases in the region associated with a low entropy for \gls{mpi}.
Looking at Fig.~\ref{fig:eve_eff_control}, we see that this phenomenon also affects Eve's accuracy and makes communication almost fully secret for $\beta \rightarrow 2$ and $\theta \rightarrow 2^7$.
Interestingly, improvement in secrecy leads to a reduction in task reward, shown in Fig.~\ref{fig:eve_eff_control}, but in a less significant manner than in the estimation scenario.

\begin{figure}[t!]
    \subfloat[Leakage.~\label{fig:leak_heu_control}]
    {\input{./tikz_fig/control/leakage_heuristic.tex}}
    \subfloat[Total reward.~\label{fig:rew_heu_control}]
    {\input{./tikz_fig/control/reward_heuristic.tex}}\\ 
    \subfloat[Control reward.~\label{fig:bob_heu_control}]
    {\input{./tikz_fig/control/acc_bob_heuristic.tex}}
    \subfloat[Eve's accuracy.~\label{fig:eve_heu_control}]
    {\input{./tikz_fig/control/acc_eve_heuristic.tex}}
    \caption{\gls{ade} performance as a function of $\theta$ and $\beta$ in the control scenario, with $D=5$.}
    \label{fig:ade_heatmaps_control}
\end{figure}

\begin{figure}[t!]
    \subfloat[Leakage.~\label{fig:leak_pde_control}]
    {\input{./tikz_fig/control/leakage_pde.tex}}
    \subfloat[Total reward.~\label{fig:rew_pde_control}]
    {\input{./tikz_fig/control/reward_pde.tex}}\\ 
    \subfloat[Control reward.~\label{fig:bob_pde_control}]
    {\input{./tikz_fig/control/acc_bob_pde.tex}}
    \subfloat[Eve's accuracy.~\label{fig:eve_pde_control}]
    {\input{./tikz_fig/control/acc_eve_pde.tex}}
    \caption{\gls{pde} performance as a function of $\theta$ and $\beta$ in the control scenario, with $D=5$.}
    \label{fig:pde_heatmaps_control}
\end{figure}

As we can observe from Fig.~\ref{fig:ade_heatmaps_control}, \gls{ade} significantly reduces the accuracy of Eve's estimates, especially in the case of a high transmission rate.
A similar effect is achieved using the \gls{pde} strategy, whose performance is instead shown in Fig.~\ref{fig:pde_heatmaps_control}.
Since its goal is to avoid $L_E$ exceeding $L_{\text{high}}$, \gls{ade} continues to use \gls{mpi} in many scenarios, especially when $\beta \rightarrow 2$ and $\theta \rightarrow 2^7$.
Instead, \gls{pde} decreases the entropy of the scheduling policy in all configurations, modifying the leakage more widely. 
As shown in Fig.~\ref{fig:leak_pde_control}, \gls{pde} obtains a lower leakage than \gls{ade}, except in scenarios with a low value of both $\theta$ and $\beta$, in which the initial \gls{mpi} scheduling is more vulnerable. 

\begin{figure}[t!]
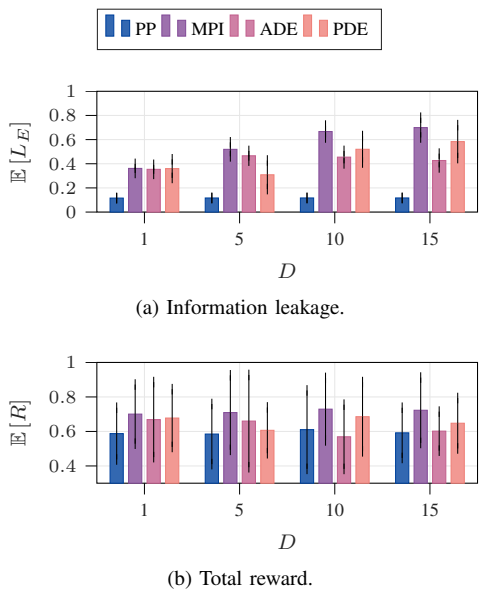

    \centering
    \subfloat{\input{tikz_fig/control/delay_legend}}\\
    \setcounter{subfigure}{0}
    \subfloat[Information leakage.~\label{fig:delay_leak_control}]
    {\input{./tikz_fig/control/delay_leakage.tex}}\\
    \subfloat[Total reward.~\label{fig:delay_rew_control}]
    {\input{./tikz_fig/control/delay_reward.tex}}
    \caption{Expected leakage and reward as a function of the opacity time gap $D$ in the control scenario, with $\beta=1$ and $\theta=32$.}
    \label{fig:delay_impact_control}
\end{figure}

In Fig.~\ref{fig:delay_impact_control}, we focus on the scenario with $\beta=1$ and $\theta=32$ and analyze the impact of the time gap $D$ on all the proposed strategies.
Fig.~\ref{fig:delay_leak_control} shows that the optimal \gls{goc} scheduling presents an average leakage below $0.6$ for $D=5$, only slightly higher than the one obtained with \gls{ade}. 
In addition, \gls{mpi} becomes more vulnerable as $D$ increases, while the average leakage of \gls{ade} never exceeds $L_{\text{high}}=0.6$.
The \gls{pde} heuristic proves to be more robust than \gls{ade} for $D\leq5$, while leaking more information as the time gap grows. 
If we consider the total reward, reported in Fig.~\ref{fig:delay_rew_control}, the relationship between the two heuristics is inverted: \gls{pde} has a degraded performance for $D\leq5$, while it constitutes an intermediate solution between \gls{mpi} and \gls{ade} for longer time gaps. 

\subsection{Pareto Analysis}

In the previous analysis, we considered specific hyperparameters for both \gls{ade} and \gls{pde}, which correspond to a single operation point.
In the following, we study the trade-off between secrecy and reward for the two heuristics by computing the Pareto frontier given by all the possible algorithm settings.
From a practical perspective, we vary the leakage thresholds $L_{\text{low}}\in[0.1, 0.7]$ while setting $L_{\text{high}} = L_{\text{low}} + 0.2$ in the case of \gls{ade}, and the target entropy $H^*\in\left[0, H(\sigma_{\text{MPI}})\right]$ in the case of \gls{pde}.
Importantly, to obtain reliable results, we run a total of $N_{\text{ep}}=50$ independent episodes per configuration.

\begin{figure}[t!]
    \centering
    \input{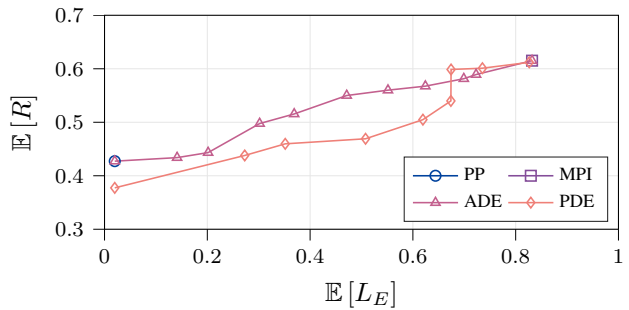}
    \caption{Pareto frontier of the trade-off between information leakage and reward in the estimation scenario, with $\beta=1$, $\theta=32$, and $D=5$.}
 \label{fig:pareto_estimation_estimation}
\end{figure}

Fig.~\ref{fig:pareto_estimation_estimation} focuses on the remote estimation case: the results show that \gls{ade} always outperforms \gls{pde} when the leakage is lower than $\sim0.7$. 
This is due to the iterative nature of \gls{pde}: suboptimal choices in the early stages of the algorithm significantly degrade performance for all the following steps.
This phenomenon is reflected by the steep performance drop experienced by \gls{pde}, which never recovers and is always outperformed by \gls{ade}. 
On the whole, \gls{ade} is able to better control the trade-off between secrecy and efficiency in this task, providing an almost linear degradation of the reward as we increase the probability of using \gls{pp}.

\begin{figure}[t!]
    \centering
    \input{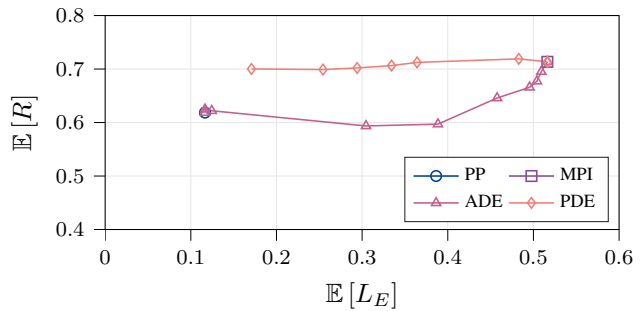}
    \caption{Pareto frontier of the trade-off between information leakage and reward in the control scenario, with $\beta=1$, $\theta=32$, and $D=5$.}
 \label{fig:pareto_control}
\end{figure}

Fig.~\ref{fig:pareto_control} repeats the analysis for the control scenario.  
In this case, \gls{pde} finds a working trajectory that allows Bob to maintain a reward very close to the optimum while strongly reducing the leakage from $0.5$ to values lower than $0.2$.
On the other hand, \gls{ade} seems inefficient in managing the control system and immediately degrades the expected reward. 
The benefits of \gls{pde} are even more relevant as, while being more sophisticated than \gls{ade} in its basic mechanism, it does not require Bob to compute the leakage online, thus greatly reducing the computational burden for real-time operation.

The remote estimation task analyzed in Fig.~\ref{fig:pareto_estimation_estimation} is characterized by monotonic relations between transition stochasticity, communication cost, and total reward. 
These trivial performance trends are unlikely in real-world applications, which are expected to be more similar to the remote control task shown in Fig.~\ref{fig:pareto_control}.
In scenarios with an irregular relationship between secrecy and efficiency, \gls{pde} is much more likely to find solutions that reduce information leakage while preserving the same performance of \gls{goc}. 
In particular, \gls{pde} makes suboptimal but more opaque scheduling decisions when this is less critical for the control reward, e.g., when Bob is farther from the target state.
Hence, the fact that the control policy is mutually adapted to communication ensure that the system experiences only a negligible performance loss.

\section{Conclusion and Future Work}
\label{sec:conc}

In this work, we analyzed the security of \gls{goc} systems for the remote control of Markov processes, focusing on the system's vulnerability to timing side-channel attacks.
This type of attack is viable even under information-theoretic secrecy, as they only rely on the presence of a message rather than its content.
We considered two different tasks, i.e., a remote estimation and a remote control problem, and analyzed four different transmission scheduling protocols: the optimal \gls{goc} scheduling, a periodic transmission policy, and two heuristic solutions that trade off between the previous approaches. 
 
Our results proved that goal-oriented scheduling has significant performance benefits, but is also highly vulnerable to eavesdropping.
In addition, although heuristic mitigation strategies are possible, finding an optimal policy under game-theoretic rationality is a computationally hard problem.
We showed that any strategy must be tuned according to the target environment, as the structure of the communication policy may vary significantly depending on factors such as the stochasticity of the system and the transmission cost.

As our study is the first to analyze timing attacks against \gls{goc}, there are many possible avenues for future work.
First, expanding the game-theoretic model may lead to more efficient heuristics.
It will be interesting to consider reinforcement learning solutions, which have properties similar to the proposed algorithms and can be deployed in more complex real-world applications.
Finally, our framework could be applied to push-based scenarios in which Alice independently decides when to send an update, which represents another attractive possibility for future research.

\bibliographystyle{IEEEtran}
\bibliography{biblio.bib}

\end{document}